\documentclass[10pt]{article}

\usepackage[margin=0.78in]{geometry}
\usepackage[T1]{fontenc}
\usepackage{lmodern}
\usepackage{amsmath,amssymb,amsfonts,mathtools}
\usepackage{booktabs,array}
\usepackage{enumitem}
\usepackage{hyperref}
\usepackage{microtype}
\usepackage{xcolor}
\usepackage{algorithm}
\usepackage{algpseudocode}

\hypersetup{
  colorlinks=true,
  linkcolor=blue,
  citecolor=blue,
  urlcolor=blue
}

\title{
Beyond Foreign-Key Connectivity:
Semantic Relationship Representation
for Relational Deep Learning
}
\author{Noy Cohen-Shapira}
\date{}

\begin{document}
\maketitle

\begin{abstract}

Relational databases encode information through tables
and primary--foreign key constraints. Although recent
Relational Deep Learning (RDL) approaches represent
these databases as heterogeneous graphs, a foreign-key
connection primarily specifies structural connectivity
between records. Its semantic interpretation may
involve additional information that is not explicitly
represented by connectivity alone.

This paper introduces a conceptual framework for
constructing relationship-aware representations from
relational databases. The framework distinguishes
three complementary aspects: (i) schema-defined
relationship semantics, including endpoint types and
foreign-key roles; (ii) association entities and
explicitly available relationship-level attributes;
and (iii) contextual information describing the
temporal and structural circumstances of a connection.

We formulate a Relationship Representation
Construction (RRC) protocol that maps relational
schema elements and database records to node
representations, relationship descriptors, and
contextual representations. A key principle of the
protocol is semantic ownership: attributes are assigned
to records, relationships, or context according to
their meaning, rather than merely their physical
location in a database table. Ambiguous attributes
are not automatically promoted to relationship
features.

The resulting representations can be incorporated
into existing relational graph learning architectures
through relationship-conditioned message passing.
The proposed framework is complementary to recent
work on relational message passing, graph transformers,
and task-dependent graph construction. We discuss its
conceptual foundations, illustrate the construction
protocol using a multi-table database, and identify
research questions and empirical evaluations required
to establish its potential benefits.

\end{abstract}

\section{Introduction}

Relational databases are the foundation of many
information systems. They organize information into
tables, where records are connected through primary-key
and foreign-key constraints. While conventional
machine learning approaches typically operate on a
single flattened table, real-world prediction tasks
often depend on information distributed across
multiple related tables.

Relational Deep Learning (RDL) addresses this
limitation by representing relational databases as
heterogeneous temporal graphs. In this representation,
each row corresponds to a graph node and
primary--foreign key relationships induce edges
between nodes. Graph neural networks can then
propagate information across tables and learn
representations that jointly capture record content
and relational structure \cite{fey2024rdl}.

Recent work has extended this paradigm in several
directions. RelGNN introduces atomic routes and
composite message passing to better exploit the
structural properties of relational graphs
\cite{chen2025relgnn}. Relational Graph Transformer
(RelGT) introduces a graph-transformer architecture
for relational data, incorporating multiple aspects
of node and structural information
\cite{dwivedi2026relgt}. Other work investigates
how the graph structure itself should be adapted
to downstream tasks through structural filtering
and injection \cite{cheng2026desired}.

These developments demonstrate that both relational
structure and the way information is propagated
through it are central to relational learning.
A complementary modeling perspective is to examine
the semantics of the relationships themselves.

A foreign-key link is not merely an indication that
two records are connected. It has a schema-defined
role determined by the database design. For example,
a link from a customer record to an order record
represents a different relationship type from a link
between an order and a product. Furthermore, some
relationships may have their own attributes, while
others may be associated with temporal or structural
context.

An important challenge is that relational database
attributes do not all have the same semantic
ownership. Some describe individual records, some
describe relationship instances, and others describe
the context in which relationships occur. Treating
all available attributes uniformly as node or edge
features may therefore obscure the meaning of the
underlying data.

This paper proposes a conceptual framework for
organizing and encoding these different forms of
relationship information. The central premise is
that relationship representations should reflect
the semantics and context of database relationships,
rather than treating all links solely as structural
connections.

The paper makes the following contributions:

\begin{itemize}[leftmargin=*]

\item We introduce a taxonomy of relationship
information in relational databases, distinguishing
schema-level semantics, instance-level attributes,
and contextual information.

\item We formulate a Relationship Representation
Construction protocol that maps relational schema
elements and record attributes to node, relationship,
and context representations according to their
semantic ownership.

\item We distinguish schema-defined links,
association entities, and explicit relationship
attributes, clarifying when an association record
should remain a node or may be represented as
an attributed edge.

\item We describe how the resulting representations
can condition graph neural network message passing.

\item We identify research questions and an empirical
evaluation agenda for investigating the potential
benefits of relationship-aware representations.

\end{itemize}

\section{Background and Related Work}

\subsection{Relational Deep Learning}

Fey et al. introduced Relational Deep Learning as
a framework for learning directly from relational
databases by converting them into heterogeneous
temporal graphs \cite{fey2024rdl}. Each database
row is represented as a node, while primary--foreign
key constraints define links between records.
This representation enables graph neural networks
to aggregate information across tables without
requiring manual flattening and feature engineering.

The resulting graph preserves the relational
structure of the database. However, translating
a database into a graph also requires decisions
about how schema semantics and relationship-specific
information should be represented.

\subsection{Relational Message Passing}

RelGNN addresses limitations of conventional
message passing on relational graphs by introducing
atomic routes and composite message passing
\cite{chen2025relgnn}. Its design enables information
exchange through relational paths, particularly
in settings involving complex relational dependencies.

RelGNN also highlights an important distinction:
foreign-key links primarily encode structural
connectivity between records and do not necessarily
correspond to semantically complete relationships.
The framework developed here is complementary
to this line of work. Whereas RelGNN emphasizes
the structural organization of information
propagation, we focus on the semantic decomposition
of information associated with database relationships
and its mapping to graph representations.

\subsection{Relational Graph Transformers}

Relational Graph Transformer introduces a transformer
architecture specifically designed for relational
data \cite{dwivedi2026relgt}. Its representation
incorporates multiple dimensions of relational
information, including node features, node types,
and structural or temporal context.

This work demonstrates the importance of representing
multiple dimensions of relational data. The
perspective developed here places particular
emphasis on the relationship as a semantic object:
its schema-defined role, any attributes that belong
to a specific relationship instance, and its
contextual properties.

This distinction is intended as a complementary
modeling perspective, rather than a claim that
existing relational architectures cannot encode
edge or structural information.

\subsection{Edge-Conditioned Graph Learning}

The general idea of conditioning graph neural
network operations on edge information predates
RDL. For example, Simonovsky and Komodakis
introduced Edge-Conditioned Convolution, in which
filter weights are conditioned on edge labels
\cite{simonovsky2017ecc}.

Accordingly, the use of edge information to
influence message passing is not itself the
novelty claimed in this paper. Our focus is on
how relationship information should be interpreted
and organized in the specific setting of relational
databases, where the semantics of a link are partly
determined by the underlying schema.

\subsection{Relational Graph Construction}

Recent work has also investigated whether graphs
derived directly from database schemas are optimal
for downstream learning. Cheng and Luo study
task-dependent graph construction through structural
filtering and injection \cite{cheng2026desired}.

It is useful to distinguish two related questions:

\begin{enumerate}[leftmargin=*]
\item \textbf{Graph construction:} Which nodes and
connections should be included in the graph?

\item \textbf{Relationship representation:} Given
a connection, what information should describe it,
and how should this information influence message
passing?
\end{enumerate}

The present framework primarily addresses the
second question. It can therefore be considered
complementary to approaches that optimize graph
structure.

\section{Problem Formulation}

\subsection{Relational Database Representation}

Let a relational database be defined as

\[
\mathcal{D} = (\mathcal{T}, \mathcal{C}, \mathcal{X}),
\]

where $\mathcal{T}$ is the set of database tables,
$\mathcal{C}$ is the set of integrity constraints,
including primary-key and foreign-key constraints,
and $\mathcal{X}$ contains the records and their
associated attribute values.

A graph representation of the database is defined as

\[
G=(V,E),
\]

where each row in a table corresponds to a node
$v\in V$, and each foreign-key reference induces
a directed relationship edge $e=(u,v)\in E$.

Throughout this paper, the default edge direction
is from the referencing record to the referenced
record. An implementation may choose the reverse
convention, provided that it is used consistently
and recorded in the schema metadata.

Each node is associated with an initial representation

\[
h_v^{(0)} = f_{\mathrm{node}}(x_v),
\]

where $x_v$ contains the attributes assigned to
the corresponding database record.

The central question addressed in this work is
how to construct a representation for each
relationship that preserves its semantic meaning
and relevant context.

\subsection{Semantic Ownership of Database Information}

Before constructing relationship representations,
the framework distinguishes three possible ownership
categories for database information:

\begin{enumerate}[leftmargin=*]

\item \textbf{Record-owned information:} attributes
describing an individual database record.

\item \textbf{Relationship-owned information:}
attributes describing a particular relationship
instance.

\item \textbf{Contextual information:} information
describing the temporal, directional, or structural
context in which a relationship is observed.

\end{enumerate}

This distinction is important because the physical
location of an attribute in a relational schema
does not necessarily determine its semantic role.

For example, an attribute such as
\texttt{order\_amount} normally describes an order
record, even if the order is connected to a customer
through a foreign key. It should therefore be
represented as a node feature of the order, rather
than automatically being assigned to the
customer--order edge.

Conversely, an attribute describing the duration
or strength of a specific interaction may be
naturally associated with the relationship instance
itself.

The framework therefore treats semantic ownership
as an explicit modeling decision.

\subsection{Distinguishing Links, Association Entities,
and Relationship Attributes}

A central distinction in the proposed framework
is between a schema-defined link and a semantic
relationship instance.

A foreign-key constraint establishes a structural
connection between two table types. At the instance
level, each valid foreign-key reference induces
a link between two corresponding records. However,
the existence of such a link does not imply that
all information associated with the connected
records should be interpreted as relationship
information.

We distinguish three cases.

\paragraph{Schema-defined links.}

A schema-defined link is induced by a foreign-key
constraint. It is characterized by its source table,
destination table, foreign-key identity, and
direction. These properties define the structural
and schema-level representation of the link.

\paragraph{Association entities.}

An association entity is a database record that
connects two or more entities through foreign keys.
Such records frequently occur in junction tables
used to represent many-to-many relationships.

An association entity may contain attributes of
its own and may participate in additional
relationships. Consequently, it should generally
remain a graph node unless there is a justified
semantic reason to collapse it into a direct
attributed edge.

\paragraph{Explicit relationship attributes.}

An explicit relationship attribute describes a
property of a particular relationship instance.
Such information may originate from an explicitly
defined relationship record, external relationship
metadata, or a schema transformation that preserves
the semantics of an association.

Importantly, attributes belonging to an endpoint
record must not be automatically interpreted as
relationship attributes.

This distinction prevents a common modeling
ambiguity: the physical representation of a
relationship in a relational database does not
necessarily determine the semantic ownership of
its attributes.

\subsection{Attribute Ownership and Uncertainty}

Let $\mathcal{A}$ denote the set of database
attributes. We define an ownership function

\[
\Omega: \mathcal{A} \rightarrow
\{
\textsc{Node},
\textsc{Relationship},
\textsc{Context},
\textsc{Schema},
\textsc{Unknown}
\}.
\]

The ownership assignment is based on schema
constraints, explicit metadata, and domain-level
semantic declarations.

The framework follows three principles:

\begin{enumerate}[leftmargin=*]

\item Attributes describing an individual record
are assigned to the corresponding node.

\item Attributes explicitly describing a
relationship instance are assigned to the
relationship representation.

\item Attributes whose ownership cannot be
established reliably are marked as unknown and
are not automatically promoted to relationship
features.

\end{enumerate}

This conservative assignment policy is important
because incorrectly transferring endpoint attributes
to edges may introduce redundant information or
alter the semantics of the original database.

The ownership function can be manually specified
initially and later extended with automatic schema
interpretation mechanisms.

\section{Relationship Representation Construction}

This section introduces the central component of
the framework: a systematic protocol for constructing
relationship representations from relational
databases.

\subsection{Three Dimensions of Relationship Information}

For each relationship edge $e=(u,v)$, we
distinguish three components.

\subsubsection{Schema-Level Relationship Semantics}

This component describes the semantic role of
a relationship according to the database schema.
It may include:

\begin{itemize}[leftmargin=*]
\item source and destination table types;
\item the identity of the foreign-key constraint;
\item the semantic role of the relationship;
\item the direction of the reference.
\end{itemize}

For example, a relationship from a
\texttt{Customer} record to an \texttt{Order}
record has a different semantic role from a
relationship connecting an \texttt{Order} to an
\texttt{OrderItem}.

We denote the schema-level representation by

\[
r_e =
f_{\mathrm{schema}}
\left(
\tau_u,\tau_v,\rho_e
\right),
\]

where $\tau_u$ and $\tau_v$ denote the endpoint
table types and $\rho_e$ denotes the relationship
role or foreign-key identity.

The schema-level representation is shared by
edges instantiating the same relationship type,
although it may also incorporate additional
schema metadata.

\subsubsection{Instance-Level Relationship Attributes}

Some relationships have properties that belong
to a particular relationship instance rather than
to either endpoint record. Examples may include
interaction strength, relationship duration,
assignment role, relationship-specific status,
and event-specific measurements.

Let $a_e$ denote the attributes associated with
a relationship instance:

\[
a_e =
f_{\mathrm{attr}}
\left(
x_e^{\mathrm{rel}}
\right).
\]

These attributes should be included only when
their semantic ownership is genuinely associated
with the relationship.

An important special case is an associative or
junction table. Such a table may represent a
relationship between two entities while also
containing additional attributes. Depending on
its semantics, the table can either be represented
as a separate node or be collapsed into an
attributed edge.

The decision should depend on whether the
association record has an independent identity,
participates in additional relationships, or
carries information that should be preserved
as a separate entity.

\subsubsection{Contextual Relationship Information}

The third component describes the context in
which a relationship exists. It may include:

\begin{itemize}[leftmargin=*]
\item temporal validity or event time;
\item directionality and traversal context;
\item cardinality-related information;
\item structural context, such as path or
neighborhood information.
\end{itemize}

We denote the contextual representation by

\[
c_e =
f_{\mathrm{context}}(q_e),
\]

where $q_e$ contains the contextual information
available for the relationship.

Not all contextual fields are necessarily edge
attributes. For example, an event timestamp may
belong to an event node. The representation
should preserve the semantic ownership of each
field and avoid duplicating information without
a clear modeling motivation.

\subsection{Formal Relationship Representation}

The three components are combined into a
relationship representation:

\[
z_e =
g_{\theta}
\left(
r_e,a_e,c_e,m_e
\right),
\]

where $g_{\theta}$ is a learnable encoder and
$m_e$ denotes optional availability indicators
or missingness information.

The availability indicators allow the framework
to handle relationships for which some components
are not available. For example, a foreign-key
relationship may have a well-defined schema-level
role but no explicit instance-level attributes.

A possible implementation is to encode each
component separately and combine the resulting
embeddings:

\[
\begin{aligned}
z_e^{\mathrm{schema}} &=
g_{\mathrm{schema}}(r_e),\\
z_e^{\mathrm{attr}} &=
g_{\mathrm{attr}}(a_e),\\
z_e^{\mathrm{context}} &=
g_{\mathrm{context}}(c_e),\\
z_e &=
g_{\mathrm{combine}}
\left(
z_e^{\mathrm{schema}},
z_e^{\mathrm{attr}},
z_e^{\mathrm{context}},
m_e
\right).
\end{aligned}
\]

This decomposition is one possible realization
of the framework. Alternative encoders and
combination mechanisms can be investigated
empirically.

\subsection{Relationship Representation Construction Protocol}

The proposed protocol consists of four main stages:

\begin{enumerate}[leftmargin=*]

\item \textbf{Schema analysis:} identify tables,
primary keys, foreign keys, and relationship types.

\item \textbf{Semantic ownership analysis:} determine
whether each attribute describes a record, a
relationship instance, or its context.

\item \textbf{Representation assignment:} construct
node features, schema-level relationship descriptors,
instance-level relationship attributes, and
contextual descriptors.

\item \textbf{Relationship encoding:} encode the
available components into a unified relationship
representation.

\end{enumerate}

The protocol is summarized in
Algorithm~\ref{alg:construction}.

\begin{algorithm}[t]
\caption{Relationship Representation Construction}
\label{alg:construction}

\begin{algorithmic}[1]

\Require Relational database $\mathcal{D}$,
schema metadata $\mathcal{S}$,
attribute ownership rules $\mathcal{O}$

\Ensure Graph $G=(V,E)$ with node and relationship
representations

\State Initialize $V \gets \emptyset$,
$E \gets \emptyset$

\State Analyze tables, keys, and foreign-key
constraints in $\mathcal{S}$

\ForAll{tables $T \in \mathcal{D}$}
    \ForAll{records $x \in T$}
        \State Identify record-owned attributes
        $x^{\mathrm{node}}$
        \State Construct node $v$ with
        $h_v^{(0)} \gets
        f_{\mathrm{node}}(x^{\mathrm{node}})$
        \State Add $v$ to $V$
    \EndFor
\EndFor

\ForAll{foreign-key relationship types $\rho$}
    \State Identify source and destination table types
    \State Construct schema descriptor $r_{\rho}$
\EndFor

\ForAll{relationship instances $e=(u,v)$}
    \State Assign schema-level information
    $r_e \gets r_{\rho(e)}$
    \State Identify relationship-owned attributes
    $a_e$ using $\mathcal{O}$
    \State Identify available contextual information
    $c_e$
    \State Construct availability indicators $m_e$
    \State Encode relationship:
    \[
    z_e \gets
    g_{\theta}(r_e,a_e,c_e,m_e)
    \]
    \State Add $(u,v,z_e)$ to $E$
\EndFor

\State \Return $G=(V,E)$

\end{algorithmic}
\end{algorithm}

The protocol separates the identification of
information from its neural encoding. This
distinction is important: the semantic assignment
of an attribute should be determined by the
database structure and its meaning, rather than
being left entirely to the neural architecture.

The protocol does not prescribe a single encoding
function. Instead, it defines the information
that a relationship-aware model should be able
to receive.

\subsection{Association Entity Transformation}

For an association table $A$ connecting entity
tables $T_1$ and $T_2$, we distinguish two
representations.

In the default node-based representation:

\[
T_1 \leftarrow A \rightarrow T_2.
\]

Each record in $A$ is represented as a node,
and its attributes remain node features.

Alternatively, when the association record has
no independent identity or additional relational
role, it may be transformed into an attributed
edge:

\[
e=(v_1,v_2), \qquad
z_e=g_\theta(x_A,r_e,c_e).
\]

This transformation should preserve the
association's attributes and multiplicity. It
should not be applied when doing so would discard
meaningful entity identity or relationships to
additional tables.

\subsection{Relationship-Conditioned Message Passing}

Given node representations $h_u^{(l)}$ and
$h_v^{(l)}$, the message transmitted from node
$u$ to node $v$ through relationship $e=(u,v)$
is defined as

\[
m_{u\rightarrow v}^{(l)}
=
\phi_{\theta}^{(l)}
\left(
h_u^{(l)},
h_v^{(l)},
z_e
\right).
\]

Messages are aggregated over incoming relationships:

\[
M_v^{(l)}
=
\operatorname{AGG}
\left(
\left\{
m_{u\rightarrow v}^{(l)}
:
(u,v)\in E
\right\}
\right).
\]

The node representation is then updated:

\[
h_v^{(l+1)}
=
\psi_{\theta}^{(l)}
\left(
h_v^{(l)},M_v^{(l)}
\right).
\]

The functions $\phi_{\theta}^{(l)}$ and
$\psi_{\theta}^{(l)}$ may be implemented using
existing edge-conditioned or relation-aware
neural operators.

The framework therefore does not require a new
general-purpose message-passing operator. Its
focus is the construction of the relationship
representation $z_e$ and the systematic use of
its semantic components during message generation.

\section{Illustrative Example}

Consider a simplified e-commerce database consisting
of the following tables:

\begin{itemize}[leftmargin=*]

\item \texttt{Customer}(
\texttt{customer\_id},
\texttt{segment},
\texttt{region})

\item \texttt{Order}(
\texttt{order\_id},
\texttt{customer\_id},
\texttt{order\_date},
\texttt{total\_amount},
\texttt{status})

\item \texttt{Product}(
\texttt{product\_id},
\texttt{category},
\texttt{price})

\item \texttt{OrderItem}(
\texttt{order\_id},
\texttt{product\_id},
\texttt{quantity},
\texttt{unit\_price})

\end{itemize}

The foreign-key constraints induce the following
relationships:

\[
\texttt{Customer}\rightarrow\texttt{Order},
\]

\[
\texttt{Order}\rightarrow\texttt{OrderItem},
\]

\[
\texttt{Product}\rightarrow\texttt{OrderItem}.
\]

The resulting graph contains nodes representing
rows from all four tables.

\subsection{Applying the Construction Protocol}

\paragraph{Stage 1: Schema analysis.}

The database schema defines three relationship
types: customer-to-order, order-to-order-item,
and product-to-order-item. These relationships
are semantically distinct even though all are
implemented using foreign-key constraints.

\paragraph{Stage 2: Semantic ownership analysis.}

The attributes are assigned according to their
semantic ownership:

\begin{itemize}[leftmargin=*]

\item \texttt{Customer.segment} and
\texttt{Customer.region} describe customer records.

\item \texttt{Order.total\_amount},
\texttt{Order.status}, and
\texttt{Order.order\_date} describe order records.

\item \texttt{Product.category} and
\texttt{Product.price} describe product records.

\item \texttt{OrderItem.quantity} and
\texttt{OrderItem.unit\_price} describe
order-item records.

\end{itemize}

In particular, \texttt{quantity} and
\texttt{unit\_price} should not automatically
be assigned to either of the foreign-key edges
connecting an order item to an order or a product.

\paragraph{Stage 3: Representation assignment.}

The three relationship types receive different
schema-level descriptors:

\[
r_1 =
f_{\mathrm{schema}}
(\texttt{Customer},\texttt{Order},
\texttt{places}),
\]

\[
r_2 =
f_{\mathrm{schema}}
(\texttt{Order},\texttt{OrderItem},
\texttt{contains}),
\]

\[
r_3 =
f_{\mathrm{schema}}
(\texttt{Product},\texttt{OrderItem},
\texttt{referenced\_by}).
\]

Temporal context may be derived from
\texttt{Order.order\_date}, but the date remains
an attribute of the order record unless the
application defines a separate relationship-level
temporal meaning.

\paragraph{Stage 4: Relationship encoding.}

For each relationship instance, the available
schema-level and contextual information is encoded
into $z_e$. If no relationship-owned attributes
are available, the instance-level component is
represented as unavailable rather than being
populated with unrelated endpoint attributes.

\subsection{Associative Tables and Relationship Attributes}

The \texttt{OrderItem} table illustrates a
particularly important modeling decision.

It connects orders and products and contains
additional attributes, such as quantity and unit
price. There are two possible representations:

\begin{enumerate}[leftmargin=*]

\item \textbf{Association-node representation:}
each \texttt{OrderItem} row is represented as a
separate node connected to its corresponding
order and product nodes.

\item \textbf{Attributed-edge representation:}
the association record is collapsed into a direct
order--product edge, with quantity and unit price
encoded as edge attributes.

\end{enumerate}

The first representation is generally more
appropriate when the association record has an
independent identity, participates in additional
relationships, or needs to be preserved as a
separate entity. The second may be appropriate
when the association is naturally interpreted
as a direct relationship and its attributes
belong to that relationship instance.

The framework does not impose one universal
choice. Instead, it makes this decision an
explicit part of relationship representation
construction.

\begin{table}[t]
\centering
\small
\caption{Illustrative mapping of relational
information to graph representations.}
\label{tab:mapping}

\begin{tabular}{p{0.25\linewidth}
                p{0.28\linewidth}
                p{0.35\linewidth}}
\toprule
\textbf{Database element} &
\textbf{Representation} &
\textbf{Reason} \\
\midrule

Customer.segment &
Node feature &
Describes a customer record. \\

Order.total\_amount &
Node feature &
Describes an order record. \\

Product.category &
Node feature &
Describes a product record. \\

OrderItem.quantity &
Association-node feature or edge attribute &
Depends on whether OrderItem is retained as a
node or collapsed into a relationship. \\

Customer--Order FK &
Schema-level relationship descriptor &
Identifies the semantic role of the connection. \\

Order date &
Node feature and potentially temporal context &
The date belongs to the order; it may additionally
inform temporal modeling. \\

Relationship direction &
Contextual descriptor &
Defines the direction of information propagation. \\

\bottomrule
\end{tabular}
\end{table}

This example demonstrates that relationship-aware
representation learning is not equivalent to moving
all attributes from relational tables onto graph
edges. Instead, it requires an explicit interpretation
of the semantic ownership and role of each piece
of information.

\section{Relationship to Existing Architectures}

The proposed framework is related to several
existing directions in graph learning and
relational deep learning. Its intended distinction
is summarized below.

\begin{table}[t]
\centering
\small
\caption{Conceptual comparison with related approaches.}
\label{tab:comparison}

\begin{tabular}{p{0.23\linewidth}
                p{0.34\linewidth}
                p{0.34\linewidth}}
\toprule
\textbf{Approach} &
\textbf{Main emphasis} &
\textbf{Perspective of this framework} \\
\midrule

Relational Deep Learning
\cite{fey2024rdl} &
Representing relational databases as heterogeneous
graphs and learning over their structure. &
Provides the underlying relational graph setting;
we focus on the construction of semantic
relationship representations. \\

RelGNN
\cite{chen2025relgnn} &
Atomic routes and composite message passing
for relational dependencies. &
Complementary focus on the semantic decomposition
of information associated with relationships. \\

Relational Graph Transformer
\cite{dwivedi2026relgt} &
Transformer-based learning with multiple
relational and structural information components. &
Emphasizes explicit semantic ownership and
mapping of relationship information. \\

Edge-Conditioned Convolution
\cite{simonovsky2017ecc} &
Conditioning graph filters on edge information. &
Provides a possible family of operators for
using the constructed relationship representation. \\

Task-dependent graph construction
\cite{cheng2026desired} &
Adapting graph structure to downstream tasks. &
Addresses what information describes a given
relationship, rather than primarily which
connections should exist. \\

\bottomrule
\end{tabular}
\end{table}

This comparison should not be interpreted as
claiming that the existing architectures lack
the ability to encode edge attributes or contextual
information. Rather, the proposed framework
provides an explicit database-oriented protocol
for deciding which information should be assigned
to nodes, relationships, and context, and how
these components can be combined.

The distinction is therefore primarily at the
level of representation construction and semantic
organization. Whether this organization yields
measurable benefits over existing architectures
remains an empirical research question.

\section{Research Questions and Design Challenges}

The proposed framework motivates several research
questions.

\subsection{RQ1: Does explicit relationship
representation improve predictive performance?}

An empirical study should compare conventional
relational GNNs with models that explicitly encode
schema-level relationship semantics and available
contextual information. The comparison should
control for model capacity and input information
to determine whether improvements arise from
the proposed representation strategy rather than
simply from adding more features.

\subsection{RQ2: Which relationship components are
most informative?}

Ablation studies should examine the contribution
of schema-level semantics, instance-level
relationship attributes, and contextual information,
individually and in combination.

\subsection{RQ3: Can relationship representations
generalize across database schemas?}

A key challenge is to learn representations that
are not tied to specific table or column names.
The use of relationship roles, endpoint types,
and shared semantic encoders may provide a basis
for transfer across databases with different schemas.

\subsection{RQ4: How should relationship information
interact with graph structure?}

Relationship representation and graph construction
are related but distinct. Future work should
investigate whether explicit relationship information
complements structural adaptation methods, including
approaches that filter or introduce graph connections.

\subsection{RQ5: How should semantic ownership
be determined?}

The protocol assumes that attributes can be assigned
to records, relationships, or context. In practice,
this assignment may be ambiguous, especially for
association tables, event tables, and denormalized
schemas.

An important research direction is to investigate
automatic or semi-automatic methods for identifying
semantic ownership from schema metadata, constraints,
and data characteristics.

Additional challenges include heterogeneous data
types, missing relationship attributes, high-cardinality
schemas, temporal consistency, and computational
scalability.

\section{Evaluation Agenda}

The framework is conceptual and requires empirical
validation. A possible evaluation should compare
the following model configurations:

\begin{enumerate}[leftmargin=*]

\item \textbf{Structural baseline:}
a relational GNN using node features and
foreign-key connectivity.

\item \textbf{Schema-aware model:}
a model that additionally encodes relationship
types and schema-level roles.

\item \textbf{Relationship-aware model:}
a model that combines schema-level semantics
with available instance-level relationship
attributes and contextual information.

\end{enumerate}

Evaluation can use relational benchmarks containing
classification, regression, and recommendation tasks.

In addition to predictive performance, experiments
should examine:

\begin{itemize}[leftmargin=*]

\item \textbf{Ablation:} the contribution of each
relationship information component.

\item \textbf{Schema transfer:} performance when
training and evaluation involve different
database schemas.

\item \textbf{Missing information:} robustness
when relationship attributes or contextual
descriptors are unavailable.

\item \textbf{Representation choices:} the effect
of modeling associative tables as nodes versus
attributed edges.

\item \textbf{Computational efficiency:} the
additional cost of relationship encoding and
message passing.

\item \textbf{Temporal robustness:} the ability
to preserve temporal consistency and generalize
across different observation periods.

\end{itemize}

A particularly important experiment is to compare
models that receive the same underlying information
but differ in how it is organized and encoded.
Such a comparison can help isolate the effect
of semantic ownership and relationship representation
from the effect of simply increasing the input
feature set.

A further evaluation dimension concerns the
accuracy of automatic semantic mapping. If a
benchmark or expert-annotated collection provides
reference ownership labels for database attributes,
a mapping component can be evaluated using measures
such as accuracy, macro-F1, and agreement with
expert annotations. Such an evaluation would assess
the mapping protocol independently of downstream
predictive performance.

No empirical results are claimed in this paper.
The evaluation agenda is intended to guide the
subsequent development and validation of the framework.

\section{Discussion}

The proposed framework emphasizes that the
construction of a relational graph representation
involves more than identifying nodes and connecting
them through foreign-key constraints. It also
requires interpreting the meaning of the information
associated with each connection.

The distinction between record-owned,
relationship-owned, and contextual information
is particularly important for relational databases
because the same physical database structure may
support multiple semantic interpretations.

For example, an associative table may represent
a relationship between two entities, an independent
event, or an entity with its own lifecycle. The
graph representation should reflect this distinction
rather than relying exclusively on the presence
of foreign keys.

The Relationship Representation Construction
protocol provides a systematic way to organize
these decisions. It separates semantic analysis
from neural encoding, allowing different encoders
and message-passing architectures to be evaluated
within a common representation framework.

Several limitations remain. First, semantic
ownership may not always be directly recoverable
from schema metadata. Second, the framework does
not prescribe a universal method for combining
the three components. Third, its potential benefits
have not yet been established empirically. Finally,
the distinction between relationship attributes
and association-entity attributes may depend on
application semantics that cannot be inferred
from database constraints alone.

These limitations motivate future work on automatic
schema interpretation, relationship representation
learning, and controlled evaluation across diverse
relational datasets.

\section{Conclusion}

This paper presented a conceptual framework for
relationship-aware representation learning in
relational databases.

The framework distinguishes schema-level
relationship semantics, instance-level relationship
attributes, and contextual relationship information.
Building on this distinction, it introduces a
Relationship Representation Construction protocol
that maps relational database information to node
features, relationship representations, and
contextual descriptors.

The framework further distinguishes schema-defined
links, association entities, and explicit relationship
attributes, and introduces a conservative semantic
ownership policy for ambiguous database attributes.
The resulting representations can be incorporated
into existing graph neural network architectures
through relationship-conditioned message passing.

The proposed perspective complements existing
Relational Deep Learning approaches by emphasizing
the semantic interpretation and systematic
construction of relationship representations.
Importantly, it does not claim that edge-conditioned
message passing is a new general-purpose mechanism.
Instead, it provides a database-oriented organization
of relationship information and a set of research
questions for future investigation.

Further work is required to develop practical
implementations, compare alternative encoding
strategies, and evaluate the benefits of the
framework across relational datasets and
downstream tasks.


\end{document}